\documentclass[aps,showpacs,superscriptaddress,preprint]{revtex4}%
\usepackage{amsfonts}
\usepackage{amsmath}
\usepackage{amssymb}
\usepackage{graphicx}%
\providecommand{\U}[1]{\protect\rule{.1in}{.1in}}

\begin{document}
\title{Electronic, mechanical, and thermodynamic properties of americium dioxide}
\author{Yong Lu}
\affiliation{LCP, Institute of Applied Physics and Computational Mathematics, Beijing
100088, China}
\author{Yu Yang}
\affiliation{LCP, Institute of Applied Physics and Computational Mathematics, Beijing
100088, China}
\author{Fawei Zheng}
\affiliation{LCP, Institute of Applied Physics and Computational Mathematics, Beijing
100088, China}
\author{Bao-Tian Wang}
\affiliation{LCP, Institute of Applied Physics and Computational Mathematics, Beijing
100088, China}
\author{Ping Zhang}
\thanks{Author to whom correspondence should be addressed. E-mail: zhang\_ping@iapcm.ac.cn}
\affiliation{LCP, Institute of Applied Physics and Computational Mathematics, Beijing
100088, China}
\affiliation{Center for Applied Physics and Technology, Peking University, Beijing 100871, China}


\begin{abstract}
By performing density functional theory (DFT) +$U$ calculations, we
systematically study the electronic, mechanical, tensile, and thermodynamic
properties of AmO$_{2}$. The experimentally observed antiferromagnetic
insulating feature [J. Chem. Phys. 63, 3174 (1975)] is successfully
reproduced. It is found that the chemical bonding character in AmO$_{2}$ is
similar to that in PuO$_{2}$, with smaller charge transfer and stronger
covalent interactions between americium and oxygen atoms. The valence band
maximum and conduction band minimum are contributed by 2$p$-5$f$ hybridized
and 5$f$ electronic states respectively. The elastic constants and various
moduli are calculated, which show that AmO$_{2}$ is less stable against shear
forces than PuO$_{2}$. The stress-strain relationship of AmO$_{2}$ is examined
along the three low-index directions by employing the first-principles
computational tensile test method. It is found that similar to PuO$_{2}$, the
[100] and [111] directions are the strongest and weakest tensile directions,
respectively, but the theoretical tensile strengths of AmO$_{2}$ are smaller
than those of PuO$_{2}$. The phonon dispersion curves of AmO$_{2}$ are
calculated and the heat capacities as well as lattice expansion curve are
subsequently determined. The lattice thermal conductance of AmO$_{2}$ is
further evaluated and compared with attainable experiments. Our present work
integrally reveals various physical properties of AmO$_{2}$ and can be
referenced for technological applications of AmO$_{2}$ based materials.

\end{abstract}
\maketitle

\section{INTRODUCTION}

Actinide based materials possess interesting physical behaviors due to the
existence of 5$f$ electrons and have attracted extensive attentions
\cite{Savrasov01,Albers01,Hecker04,Moore06,Prodan05,Prodan06,Prodan07}, among
which the actinide dioxides (AnO$_{2}$) have been investigated most widely. At
their ground state, all actinide dioxides crystallize in face-centered-cubic
(CaF$_{2}$-like) structure. In this arrangement, each actinide atom, located
at the center of an oxygen cube, is expected in the ionic limit to yield four
of its electrons to the surrounding oxygen atoms. This would lead to formal
integer 5$f$ orbital populations ranging from $f^{0}$ (Th) to $f^{10}$ (Fm).
Among the dioxide series, PuO$_{2}$ and AmO$_{2}$ occupy the intermediate
zone, before PuO$_{2}$ (ThO$_{2}$ $\rightarrow$ NpO$_{2}$) the 5$f$ electronic
states appear as localized states above the oxygen-2$p$ states, while after
AmO$_{2}$ (CmO$_{2}$ $\rightarrow$ FmO$_{2}$) the 5$f$ electronic states
appear as localized states below the oxygen-2$p$ states. Thus the actinide
5$f$ electronic states in AmO$_{2}$ as well as in PuO$_{2}$ have the largest
overlap with oxygen-2$p$ states. However, different from the widespread
researches on PuO$_{2}$
\cite{Prodan07,Sun08JCP,Jomard08,Andersson09,Zhang10,Nakamura10}, AmO$_{2}$
have received very few theoretical concerns until now. Therefore following our
previous systematic investigations on the ground-state properties and
high-pressure behaviors of PuO$_{2}$ \cite{Zhang10}, here we further
investigate the corresponding physical properties of AmO$_{2}$.

Another reason for a thorough study of AmO$_{2}$ is its important role played
in the latest nuclear reactor fuels. During the burning cycle of UO$_{2}$ in
conventional fission nuclear reactors, some considerable amounts of plutonium
and neptunium isotopes, as well as smaller quantities of minor actinides such
as Am, Cm, Bk, and Cf, emerge in the reaction waste \cite{Petit10}. Since such
radioactive waste results in very troublesome long-term storage requirements,
it is then suggested that the produced transuranium element dioxides could be
reprocessed from the burnt fuel and used as alternative fuel in a new
generation of reactors \cite{Duriez00,Inoue00,Arima06,Martin07}. Therefore,
the mechanical and thermodynamic properties of these dioxides also need to be
investigated and presented. Similar to our previous theoretical studies on
PuO$_{2}$ \cite{Sun08JCP,Zhang10} and NpO$_{2}$ \cite{WangBT10}, here we will
systematically obtain various physical properties of AmO$_{2}$ and compare
them with attainable experiments as well as possible.

Different physical properties of AmO$_{2}$ have been investigated in
experiments ever since 1969 \cite{Kalvius69,Karraker75}. The earlier magnetic
susceptibility measurements reveal that AmO$_{2}$ undergoes antiferromagnetic
transition at 8.5$\pm$0.5 K \cite{Karraker75}. Recently, along with the
development of computing abilities and numerical methods for applying quantum
mechanics calculations, more and more theoretical researches are now being
carried out on multi-phase properties of different actinide dioxides. Although
problems still exist like on the ground-state magnetic orders
\cite{Wilkins06,Santini09,Sanati11,Petit03,Yin08}, many theoretically obtained
physical properties of dioxides like UO$_{2}$ and PuO$_{2}$ already accord
very well with experiments. For example, the optical spectrum of UO$_{2}$
given by employing the density functional theory (DFT) +$U$ method
\cite{Shi10} is well consistent with the optical reflectance measurement
\cite{Sch}, which has inspired experimentalist \cite{Bruycker10} to choose the
DFT+$U$ results of PuO$_{2}$ \cite{Shi10} as a criteria in determining the
melting temperature of this material. Also remarkably, the DFT+$U$ predicted
phonon spectrum of PuO$_{2}$ \cite{Zhang10} has been confirmed by the most
recent experimental report \cite{Manley12}. Based on these previous successes,
in the present paper we intend to reveal the ground-state properties for
AmO$_{2}$, including electronic, mechanical, and thermodynamic aspects, as
well as its high-pressure behaviors.

\section{Computation Methods}

Our DFT calculations are carried out using the Vienna \textit{ab initio}
simulations package (VASP) \cite{VASP} with the projected-augmented-wave (PAW)
potential methods \cite{PAW}. The exchange and correlation effects are
described by local density approximation (LDA) and generalized gradient
approximation (GGA) in the Perdew-Burke-Ernzerhof (PBE) form \cite{PBE}. The
plane-wave basis set is limited by an energy cutoff of 500 eV. Integrations
over the Brillouin Zone of the cubic cell containing 4 americium and 8 oxygen
atoms are done on a 9$\times$9$\times$9 $k$ points mesh generated by the
Monkhorst-Pack \cite{Monkhorst76} method, which is sufficient for an energy
convergence of less than 1.0$\times$10$^{-4}$ eV per atom. The strong on-site
Coulomb repulsion among the localized americium 5$f$ electrons is described by
the DFT+$U$ formalism formulated by Dudarev \emph{et al.}
\cite{Dudarev97,Dudarev98,Dudarev00}. In this method, only the difference
between the spherically averaged screened Coulomb energy $U$ and the exchange
energy $J$ is significant for the total LDA/GGA energy functional. And in this
paper, the Coulomb $U$ is treated as a variable, while the exchange energy is
set to be a constant $J$=0.75 eV. Since only the difference between $U$ and
$J$ is significant, thus, we will henceforth label them as one single
parameter, labeled as $U$ for simplicity, while keeping in mind that the
nonzero $J$ has been used during calculations.

In order to calculate the elastic constants for AmO$_{2}$, we enforced small
strains on the equilibrium cubic cell. For small strain $\epsilon$, Hooke's
law is valid and the crystal energy $E(V,\epsilon)$ can be expanded as a
Taylor series \cite{Nye85},
\begin{equation}
E(V,\epsilon)=E(V_{0},0)+V_{0}\sum_{i=1}^{6}\sigma_{i}\emph{e}_{i}+\frac
{V_{0}}{2}\sum_{i=1}^{6}\sum_{j=1}^{6}C_{ij}e_{i}e_{j}+O(\{e_{i}^{3}\}),
\label{eq1}%
\end{equation}
where $E(V_{0},0)$ is the total energy at the equilibrium volume \emph{V$_{0}%
$} of the cell without strains, $C_{ij}$ are the elastic constants, and
$\epsilon$ is the strain tensor which has matrix elements $\varepsilon_{ij}$
($i,j$=1, 2, and 3) defined as
\begin{equation}
\varepsilon_{ij}=\left(
\begin{array}
[c]{ccccccc}
&  &  &  &  &  & \\
e_{1} & \frac{1}{2}e_{6} & \frac{1}{2}e_{5} & \frac{1}{2}e_{6} & e_{2} &
\frac{1}{2}e_{4} & \frac{1}{2}e_{5}\\
&  &  &  &  &  & \\
\frac{1}{2}e_{6} & e_{2} & \frac{1}{2}e_{4} & \frac{1}{2}e_{5} & \frac{1}%
{2}e_{4} & e_{3} & \\
&  &  &  &  &  & \\
\frac{1}{2}e_{5} & \frac{1}{2}e_{4} & e_{3} &  &  &  & \\
&  &  &  &  &  & \\
&  &  &  &  &  &
\end{array}
\right)  . \label{eq2}%
\end{equation}
Note that we have used the Voigt notation in the equation above which replaces
$xx$, $yy$, $zz$, $yz$, $xz$, and $xy$ by 1, 2, 3, 4, 5, and 6, respectively.
For cubic structure, there are only three independent elastic constants,
$C_{11}$, $C_{12}$, and $C_{44}$. So, we can employ three different strains to
calculate them as follows, \textbf{$\epsilon$}$^{1}$=($\delta$,$\delta
$,0,0,0,0), \textbf{$\epsilon$}$^{2}$=($\delta$,$-\delta$,0,0,0,0),
\textbf{$\epsilon$}$^{3}$=(0,0,0,$\delta$,$\delta$,0). The strain amplitude
$\delta$ is varied in steps of 0.02 from $-$0.06 to 0.06 and the total
energies $E(V,\delta)$ at these strain steps are calculated.

For semiconductors, the Helmholtz free energy $F$ can be expressed as
\begin{equation}
F(V,T)=E(V)+F_{vib}(V,T), \label{eq3}%
\end{equation}
where $E(V)$ stands for the ground-state electronic energy, and $F_{vib}(V,T)$
is the phonon free energy at a given unit cell volume $V$. Under quasi-hamonic
approximation (QHA), $F_{vib}(V,T)$ can be evaluated by
\begin{equation}
F_{vib}(V,T)=k_{B}T\sum_{j,\mathbf{q}}\ln\left[  2\sinh\left(  \frac
{\hbar\omega_{j}(\mathbf{q},V)}{2k_{B}T}\right)  \right]  , \label{eq4}%
\end{equation}
where $\omega_{j}(\mathbf{q},V)$ is the phonon frequency of the $j$th phonon
mode with wave vector $\mathbf{q}$ at fixed $V$, and $k_{B}$ is the Boltzmann
constant. The total specific heat of the crystal is the sum of all phonon
modes over the Brillouin zone (BZ),
\begin{equation}
C_{v}(T)=\sum_{j,\mathbf{q}}c_{v,j}(\mathbf{q},T). \label{eq5}%
\end{equation}
$c_{v,j}(\mathbf{q},T)$ is the mode contribution to the specific heat defined
as
\begin{equation}
c_{v,j}(\mathbf{q},T)=k_{B}\sum_{j,\mathbf{q}}\left(  \frac{\hbar\omega
_{j}(\mathbf{q},V)}{2k_{B}T}\right)  ^{2}\frac{1}{\sinh^{2}[\hbar\omega
_{j}(\mathbf{q},V)/2k_{B}T]}. \label{eq6}%
\end{equation}
The mode Gr\"{u}neisen parameter $\gamma_{j}(\mathbf{q})$ describing the
phonon frequency shift with respect to the volume can be calculated by
\begin{equation}
\gamma_{j}(\mathbf{q})=-\frac{d[\ln\omega_{j}(\mathbf{q},V)]}{d[\ln V]}.
\label{eq7}%
\end{equation}

The lattice thermal conductivity $\kappa$ for a material can be written as a
sum over one longitudinal ($\kappa_{LA}$) and two transverse ($\kappa_{TA}$
and $\kappa_{TA^{\prime}}$) acoustic phonon branches \cite{Morelli02,Palmer97}%
,
\begin{equation}
\kappa=\kappa_{LA}+\kappa_{TA}+\kappa_{TA^{\prime}}. \label{eq8}%
\end{equation}
These partial thermal conductivities can be calculated differently, depending
on the specific mechanisms for phonon scattering rates $1/\tau_{c}$, where
$\tau_{c}$ is the relaxation time. At relative high temperatures, the dominant
mechanism for phonon scattering is the normal and Umklapp phonon-phonon
processes ($1/\tau_{c}=1/\tau_{N}+1/\tau_{U}$), in which mainly the acoustic
phonon branches interact with each other and transport heat. Using the
Debye-Callaway model \cite{Morelli02,Callaway59}, the partial conductivities
$\kappa_{i}$ ($i$ corresponds to TA, TA$^{\prime}$, or LA modes) can be
expressed as
\begin{equation}
\kappa_{i}=\frac{1}{3}C_{i}T^{3}\left\{  \int_{0}^{\Theta_{i}/T}\frac{\tau
_{c}^{i}(x)x^{4}e^{x}}{(e^{x}-1)^{2}}dx+\frac{\left[  \int_{0}^{\Theta_{i}%
/T}\frac{\tau_{c}^{i}(x)x^{4}e^{x}}{\tau_{N}^{i}(e^{x}-1)^{2}}dx\right]  ^{2}%
}{\int_{0}^{\Theta_{i}/T}\frac{\tau_{c}^{i}(x)x^{4}e^{x}}{\tau_{N}^{i}\tau
_{U}^{i}(e^{x}-1)^{2}}dx}\right\}  , \label{eq9}%
\end{equation}
where $\Theta_{i}$ is the longitudinal (transverse) Debye temperature,
$x=\hbar\omega/k_{B}T$, and $C_{i}=k_{B}^{4}/2\pi^{2}\hbar^{3}v_{i}$. Here,
$\hbar$ is the Plank constant, $k_{B}$ is the Boltzmann constant, $\omega$ is
the phonon frequency, and $v_{i}$ is the longitudinal or transverse acoustic
phonon velocity. The normal phonon scattering and Umklapp phonon-phonon
scattering are written as
\begin{align}
\frac{1}{\tau_{N}^{LA}(x)}  &  =\frac{k_{B}^{3}\gamma_{LA}^{2}V}{M\hbar
^{2}v_{LA}^{5}}\left(  \frac{k_{B}}{\hbar}\right)  ^{2}x^{2}T^{5},\\
\frac{1}{\tau_{N}^{TA/TA^{\prime}}(x)}  &  =\frac{k_{B}^{4}\gamma
_{TA/TA^{\prime}}^{2}V}{M\hbar^{3}v_{TA/TA^{\prime}}^{5}}\frac{k_{B}}{\hbar
}xT^{5}, \label{eq10-11}%
\end{align}
and
\begin{equation}
\frac{1}{\tau_{U}^{i}(x)}=\frac{\hbar\gamma^{2}}{Mv_{i}^{2}\Theta_{i}}\left(
\frac{k_{B}}{\hbar}\right)  ^{2}x^{2}T^{3}e^{-\Theta_{i}/3T}, \label{eq12}%
\end{equation}
respectively, where $\gamma$ is the Gr\"{u}neisen parameter, $V$ is the volume
per atom, and $M$ is the average mass of an atom in the crystal. With
reasonable expressions of the Debye temperature and acoustic Gr\"{u}neisen
parameter to describe the harmonic phonon branches and the anharmonic
interactions between different phonon branches, Eq. (\ref{eq9}) can provide
reasonable predictions for a material's thermal conductivity.

\section{Results and discussions}

\subsection{Structural and electronic properties}

\begin{table}[ptb]
\caption{Calculated lattice parameters ($a_{0}$), bulk modulus ($B_{0}$),
pressure derivative of bulk modulus ($B^{\prime}_{0}$), spin moment of each
americium ion ($\mu_{mag}$), and energy band gap ($E_{g}$) for
antiferromagnetic AmO$_{2}$ by different calculational methods. Other
experimental and theoretical results are also included for comparisons.}%
\begin{ruledtabular}
\begin{tabular}{cccccccccccccccc}
Method&$a_{0}$&$B_{0}$&$B_{0}^{\prime}$&$\mu_{mag}$&$E_{g}$\\
&(\AA)&(GPa)&&($\mu_{B}$)&(eV)\\
\hline
LDA+$U$ ($U$=0)&5.280&217.7&4.53&4.56&0\\
LDA+$U$ ($U$=4)&5.351&189.6&4.71&4.94&0.7\\
GGA+$U$ ($U$=0)&5.401&179.3&4.10&4.77&0\\
GGA+$U$ ($U$=4)&5.484&140.1&4.81&5.15&1.0\\
Expt.&5.383$^{a}$,5.375$^{b}$&\\
HSE&5.37$^{c}$&&&5.1$^{c}$&1.6$^{c}$\\
\end{tabular} \label{a0}
\\
\begin{flushleft}
$^{a}$ Reference \onlinecite{Templeton53}. \\
$^{b}$ Reference \onlinecite{Nishi08}. \\
$^{c}$ Reference \onlinecite{Prodan07}. \\
\end{flushleft}
\end{ruledtabular}
\end{table}

Firstly, the equilibrium lattice parameter of AmO$_{2}$ is determined from
total-energy calculations by the DFT+$U$ methods. The total energies are
calculated in a wide range of lattice constants and fitted to the
Brich-Murnaghan equation of states (EOS) \cite{Brich47},
\begin{align}
E(V)  &  =E_{0}+\frac{9V_{0}B_{0}}{16}\left\{  \left[  \left(  \frac{V_{0}}%
{V}\right)  ^{2/3}-1\right]  ^{3}B_{0}^{\prime}\right. \nonumber\\
&  \left.  +\left[  \left(  \frac{V_{0}}{V}\right)  ^{2/3}-1\right]
^{2}\left[  6-4\left(  \frac{V_{0}}{V}\right)  ^{2/3}\right]  \right\}  ,
\end{align}
where $E_{0}$ is the electronic energy of AmO$_{2}$ at its equilibrium volume
$V_{0}$, $B_{0}$ is the bulk modulus, and $B_{0}^{\prime}$ is the pressure
derivative of the bulk modulus. The numerically fitted lattice constants,
$B_{0}$, and $B_{0}^{\prime}$ are summarized in Table I. For comparison, the
experimental values \cite{Templeton53,Nishi08} and HSE theoretical results
\cite{Prodan07} are also listed. One can see that the LDA and LDA+$U$ results
on the lattice constant $a_{0}$ are both smaller, while the GGA and GGA+$U$
results are both larger than the experimental values. This fact reflects the
overbinding effect of LDA and the underbinding effect of GGA on the chemical
bonding strengths between different atoms. However, all the obtained lattice
constants are not too far away from the experimental values. As for bulk
modulus $B_{0}$, it displays strong dependence on the $U$ values for both
LDA+$U$ and GGA+$U$ approaches. As $U$ increases from 0 to 4 eV, $B_{0}$
decreases monotonously by 12.9\% and 21.9\% for LDA+$U$ and GGA+$U$
calculations, respectively. Due to overbinding effect of LDA and underbinding
effect of GGA, the LDA+$U$ results of $B_{0}$ are always higher than the
GGA+$U$ ones. We have also calculated the bulk modulus $B_{0}$ from empirical
formula based on the elastic constants (as shown in Table III), which turns
out to be very close to the result by EOS fitting, indicating that our
calculations are self-consistent. Moreover, through EOS fitting, we also
systematically obtain the pressure derivative of bulk modulus by using
different methods. The corresponding results are also listed in Table I.

\begin{figure}[ptb]
\includegraphics[width=0.8\textwidth]{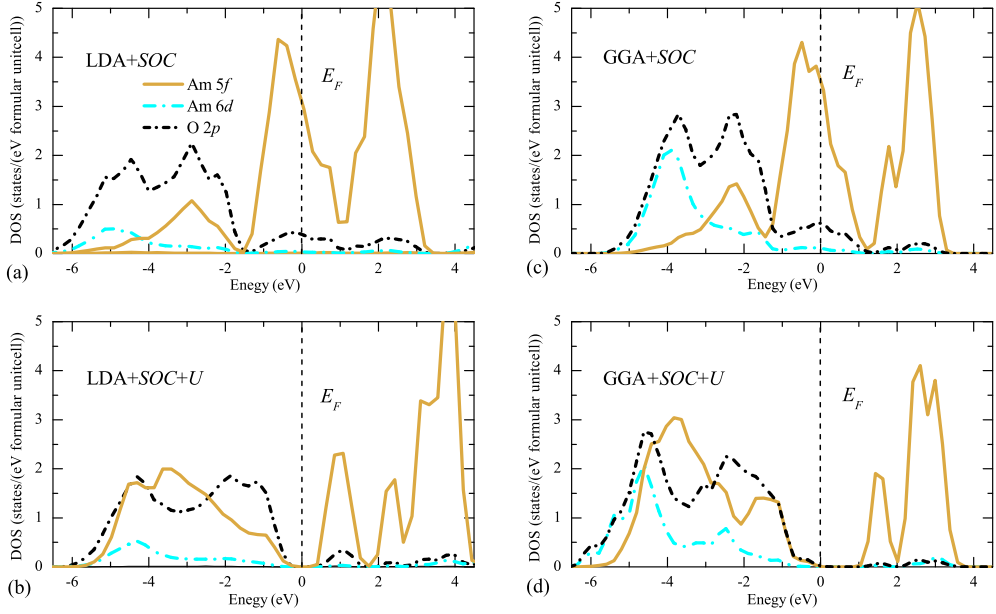}\caption{(Color online).
Electronic density of states for AmO$_{2}$ in its AFM state calculated by
using the (a) LDA, (b) LDA+$U$, (c) GGA, and (d) GGA+$U$ methods. The
spin-orbit coupling is included in all calculations. The value for the $U$
parameter is chosen to be 4 eV for all LDA+$U$ and GGA+$U$ calculations. The
Fermi energy is set at zero.}%
\label{fig:fig1}%
\end{figure}

Through electronic structure calculations, we find that without employing the
+$U$ method describing the strong electron correlation effect, AmO$_{2}$ is
wrongly described to be metallic in its ground state. Figure 1 shows the
projected density of states (PDOS) for AmO$_{2}$ obtained with different
computational methods. Since spin-orbital coupling (SOC) effects might be
important to electronic structure descriptions of actinide dioxides, here we
consider SOC in all calculations. In the LDA/GGA calculational level, SOC
makes americium 5$f$ orbitals to split into the lower $j$=5/2 ($f_{5/2}$) and
higher $j$=7/2 ($f_{7/2}$) orbitals, occupying below and above the Fermi
energy respectively, as shown in Figs. 1(a) and 1(c). The obtained electronic
states show metallic characters with considerable $f_{5/2}$ states
distributing at the Fermi energy. However, after turning on the $U$-parameter
to describe the strong on-site electron correlation effects of americium 5$f$
electrons, the $f_{5/2}$ state is further split into two separate sub-bands,
distributing at the two sides of the Fermi energy. As shown in Figs. 1(b) and
1(d), an obvious energy band gap appears in LDA/GGA+$U$ calculations. Within
LDA and GGA+$U$ calculations, the energy band gap of AmO$_{2}$ is found to be
0.7 and 1.0 eV respectively, as shown in Table I. As will be discussed below,
the difference between the obtained energy band gaps might come from the
different hybridization strengths of the americium 5$f$ and 6$d$ states in
LDA/GGA+$U$ calculations. The obtained insulating ground state of AmO$_{2}$ is
in agreement with previous hybrid density functional result \cite{Prodan07},
where the energy band gap is found to be 1.6 eV. From the obtained PDOS
distributions shown in Figs. 1(b) and 1(d), we can see that the americium 5$f$
and 6$d$ states hybridize below the Fermi energies, and this hybridization is
stronger in GGA+$U$ than in LDA+$U$ calculations. The stronger hybridization
moves the 5$f$ states lower, and results in the fact that the band gap is
larger in GGA+$U$ calculations. The calculated local spin moment of each
americium ion are 4.94 $\mu_{B}$ and 5.15 $\mu_{B}$ in the LDA+$U$ and GGA+$U$
calculations respectively, which are both similar to the 5.1 $\mu_{B}$ result
obtained in hybrid density functional calculations \cite{Prodan07}.

\begin{figure}[ptb]
\includegraphics[width=0.35\textwidth]{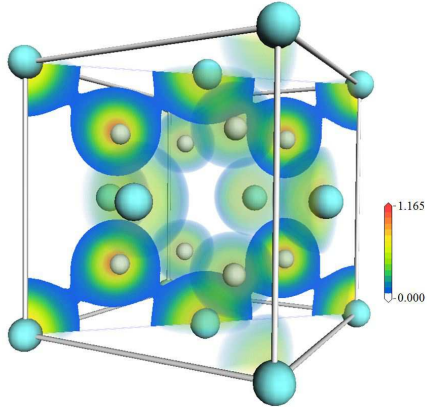}\caption{(Color online). The
cubic cell of AmO$_{2}$ with $Fm\bar{3}m$ space group, and a contour map
depicting the charge density distribution inside the (110) plane. Large cyan
and small white balls represent for americium and oxygen atoms respectively.
The charge density distribution is calculated by using the GGA+$U$ method with
the $U$ parameter of 4 eV.}%
\label{fig:fig2}%
\end{figure}

To further understand the chemical bonding characters of AmO$_{2}$, we present
in Fig. 2 the charge density distribution inside the (110) plane. Similar to
what we have found in our previous studies on other actinide dioxides
\cite{Zhang10,WangBT10}, the charge density distribution around the Am and O
ions are both spherical, with small deformations in the directions toward
their nearest neighboring ions. This character is typical for mixed
ionic/covalent chemical bondings in actinide dioxides \cite{Zhang10}. By
employing the Bader analysis \cite{Bader1,Bader2} which decomposes the charge
density distribution in real space following topological methods, we further
calculate the atomic charges for Am and O ions and present the results in
Table II. Note that although we have included the core charge in charge
density calculations, only the valence charge results are listed. Given that
the valence electron numbers of Am and O atoms are 17 and 6, about 2.26
electrons transfer from each Am to O atoms. In our previous studies, we have
also applied similar DFT+$U$ calculations for UO$_{2}$, NpO$_{2}$, and
PuO$_{2}$ \cite{Zhang10,WangBT10} and found that the charge transfer from each
actinide to oxygen atoms is 2.56 for UO$_{2}$, 2.48 for NpO$_{2}$, and 2.40
for PuO$_{2}$. From the charge transfer results, we can see that with
increasing the atomic number of the actinide elements, the chemical bondings
between actinide and oxygen atoms show less ionic characters. In addition, the
bonding lengths of the dioxides are also listed in Table II. Accompanying with
the weakening of ionicity, the bonding lengths between actinide and oxygen
atoms decrease with increasing the atomic number of actinide elements.

\begin{table}[ptb]
\caption{Calculated atomic charges according to Bader partitioning as well as
the bonding lengths of different actinide dioxides.}%
\begin{ruledtabular}
\begin{tabular}{cccccccccccccccc}
& $Q_{B}(A)$ & $Q_{B}$(O) & bonding length\\
Compound & ($e$) & ($e$) & ({\AA})\\
\hline
AmO$_{2}$   & 14.74 & 7.13 & 2.27 \\
PuO$_{2}$$^a$ & 13.60 & 7.20 & 2.32 \\
NpO$_{2}$$^b$ & 12.52 & 7.24 & 2.34 \\
UO$_{2}$$^a$  & 11.44 & 7.28 & 2.36 \\
ThO$_{2}$$^c$ &  9.34 & 7.33 & 2.43 \\
\end{tabular}\label{chg}
\\
\begin{flushleft}
$^{a}$ Reference \onlinecite{Zhang10}. \\
$^{b}$ Reference \onlinecite{WangBT10}. \\
$^{c}$ Reference \onlinecite{WangJNM10}. \\
\end{flushleft}
\end{ruledtabular}
\end{table}

\subsection{Mechanical properties}

Based on the Hooke's law, we systematically calculate the three independent
elastic constants for AmO$_{2}$ within the GGA+$U$ formalism. The results are
listed in Table III together with our previous results on other actinide
dioxide materials. At the value of $U$=4 eV, the calculated values of $C_{11}%
$, $C_{12}$, and $C_{44}$ are 250.4, 87.0, and 55.3 GPa respectively.
Mechanically, the phase of AmO$_{2}$ is stable due to the fact that its
elastic constants satisfy the following mechanical stability criteria
\cite{Nye85} of cubic structure:
\begin{equation}
C_{11}>0,~C_{44}>0,~C_{11}>|C_{12}|,~(C_{11}+2C_{12})>0. \label{eq13}%
\end{equation}
Among the actinide dioxide series listed in Table III, the elastic constants
of AmO$_{2}$ have the lowest values. Based on the elastic constants, we can
further calculate the bulk and shear moduli from the Voigt-Reuss-Hill (VRH)
approximations. Firstly, the Voigt and Reuss limits on the bulk
($B_{\mathrm{V}}$ and $B_{\mathrm{R}}$) and shear moduli ($G_{\mathrm{V}}$ and
$G_{\mathrm{R}}$) are calculated following the expressions in Refs.
\onlinecite{Voigt28,Reuss29}, then the bulk and shear moduli can be
approximately evaluated by $B=\frac{1}{2}(B_{\mathrm{V}}+B_{\mathrm{R}})$ and
$G=\frac{1}{2}(G_{\mathrm{V}}+G_{\mathrm{R}})$ \cite{Hill52}. In addition, the
Young$^{\prime}$s modulus $E$ and Poisson$^{\prime}$s ratio $\nu$ can also be
evaluated from the elastic constants by $E=\frac{9BG}{3B+G}$ and $\nu
=\frac{3B-2G}{2(3B+G)}$, respectively. Our results for mechanical properties
of AmO$_{2}$ by using the GGA+$U$ method are collected in Table III. The value
of 141.5 GPa for bulk modulus $B$ is in agreement with the EOS-fitting result
of 140.1 GPa. Besides, in comparison with other actinide dioxides included in
Table III, the value of $B$ for AmO$_{2}$ is much lower. This means that when
external pressures are applied on these actinide dioxides, AmO$_{2}$ will be
the easiest to get cracked. For the shear modulus, we can see that PuO$_{2}$
and AmO$_{2}$ have relatively smaller values than UO$_{2}$, NpO$_{2}$, and
ThO$_{2}$. This result comes from the fact that the oxygen 2$p$ and actinide
5$f$ electronic states have more overlaps in PuO$_{2}$ and AmO$_{2}$, and
subsequently the chemical bondings between oxygen and actinide atoms show more
covalent characters in PuO$_{2}$ and AmO$_{2}$. Considering that the values of
$B$ and $G$ represent for the ability to resist external pressures and
shearing forces respectively, Pugh \textit{et al.} suggested that their
quotient $B$/$G$ can be used to scale the ductility or brittleness of a solid
\cite{Pugh54}. A high value of $B$/$G$ means that the material has more
ductility than brittleness, and a critic value of 1.75 is suggested to
distinguish ductile or brittle materials. Our calculated value of 2.19 for
$B$/$G$ means that AmO$_{2}$ is a ductile material. At last, our calculated
Young's modulus and Poisson's ratio for AmO$_{2}$ are 168.4 GPa and 0.302
respectively, both of which are among the typical values of insulating oxide materials.

\begin{table}[ptb]
\caption{Calculated elastic constants, various moduli, and Poisson$^{\prime}$s
ratio $\nu$ for AmO$_{2}$. For comparisons, previous experimental values and
other theoretical results are also listed.}%
\begin{ruledtabular}
\begin{tabular}{cccccccccccccccc}
& $C_{11}$ & $C_{12}$ & $C_{44}$ & $B$  & $G$ & $E$ & $\nu$ & $v_t$ & $v_l$ & $v_m$ & $\Theta$ \\
& (GPa)    & (GPa)    & (GPa)    & (GPa)&(GPa)&(GPa)&   ~   & (m/s) & (m/s) & (m/s) & (K) \\
\hline
AmO$_{2}$                    & 250.4 &  87.0 & 55.3 & 141.5 & 64.7 & 168.4 & 0.302 & 2416.6 & 4534.1 & 2699.8 & 335.7 \\
PuO$_{2}$$^a$ & 256.5 & 167.9 & 59.2 & 197   & 52.7 & 145.2 & 0.377 & 2477.6 & 5206.3 & 2787.0 & 354.5 \\
NpO$_{2}$$^b$ & 363.6 & 118.8 & 57.4 & 200   & 78.1 & 207.5 & 0.327 & 2835.0 & 5566.5 & 3176.8 & 401.2 \\
UO$_{2}$$^a$  & 389.3 & 138.9 & 71.3 & 222.4 & 89.5 & 236.7 & 0.323 & 2841.8 & 5552.7 & 3183.4 & 398.1 \\
ThO$_{2}$$^c$ & 349.5 & 111.4 & 70.6 & 191   & 87.1 & 226.8 & 0.302 & 2969.1 & 5575.5 & 3317.3 & 402.6 \\
\end{tabular}\label{mechanical}
\\
\begin{flushleft}
$^{a}$ Reference \onlinecite{Zhang10}. \\
$^{b}$ Reference \onlinecite{WangBT10}. \\
$^{c}$ Reference \onlinecite{WangJNM10}. \\
\end{flushleft}\end{ruledtabular}
\end{table}

The Debye temperature ($\Theta$) of a solid can be determined following the
Debye theory, which describes atomic vibrations to be elastic waves, and
subsequently $\Theta$ is related to the sound velocity of a material
\cite{Blanco04}. Specifically,
\begin{equation}
\Theta=\frac{h}{k_{B}}\left(  \frac{3n}{4\pi\Omega}\right)  ^{1/3}v_{m},
\label{eq14}%
\end{equation}
where $h$ and $k_{B}$ are Planck and Boltzmann constants, respectively, $n$ is
the number of atoms in one unit cell, $\Omega$ is the volume per unit cell,
and $v_{m}$ is the averaged sound velocity. Approximately, the averaged sound
velocity can be calculated by
\begin{equation}
v_{m}=\left[  \frac{1}{3}\left(  \frac{2}{v_{t}^{3}}+\frac{1}{v_{l}^{3}%
}\right)  \right]  ^{-1/3}, \label{eq15}%
\end{equation}
where $v_{t}=\sqrt{G/\rho}$ ($\rho$ is the density of a material) and
$v_{l}=\sqrt{(3B+4G)/3\rho}$ are the transverse and longitudinal elastic wave
velocities, $B$ and $G$ are the bulk and shear moduli. Our calculated sound
velocities and Debye temperature of AmO$_{2}$ are also listed in Table III.
For comparison, our previous results for other actinide dioxides are also
included. We can see from Table III that the $v_{t}$ and $v_{l}$ values of
AmO$_{2}$ are lower than those of PuO$_{2}$, UO$_{2}$, NpO$_{2}$, and
ThO$_{2}$. Since the sound velocity of a compound is always related to its
lattice thermal conductivity, this result reflects that AmO$_{2}$ has smaller
thermal transport properties. Because the Debye temperature is proportional to
$v_{m}$, AmO$_{2}$ among the listed actinide dioxides possesses the lowest
Debye temperature.

\subsection{Theoretical tensile strength}

Based on the obtained atomic structure of AmO$_{2}$, we further study its
theoretical tensile strength. The ideal strength of materials is the stress
that is required to force deformation or fracture at the elastic instability.
Although the strength of a real crystal can be changed by the existing cracks,
dislocations, grain boundaries, and other microstructural features, its
theoretical value can never be raised, i.e., the theoretical strength sets an
upper bound on the attainable stress. Here, we employ a first-principles
computational tensile test (FPCTT) \cite{YZhang07} to calculate the
stress-strain relationship and obtain the ideal tensile strength by deforming
the AmO$_{2}$ crystals to failure. The anisotropy of the tensile strength is
tested by pulling the initial fluorite structure along the low-index [001],
[110], and [111] directions. As shown in Figs. 3(a)-3(c), three geometric
structures are constructed to investigate the tensile strengths in the three
high-symmetry crystallographic directions: Fig. 3(a) shows a general fluorite
structure of AmO$_{2}$ with four Am and eight O atoms; Fig. 3(b) a
body-centered tetragonal unit cell with two Am and four O atoms; and Fig. 3(c)
an orthorhombic unit cell with six Am and twelve O atoms. The tensile stress
is calculated according to the Nielsen-Martin scheme \cite{Nielsen85}
$\sigma_{\alpha\beta}=\frac{1}{\Omega}\frac{\partial E_{total}}{\partial
\varepsilon_{\alpha\beta}}$, where $\varepsilon_{\alpha\beta}$ is the strain
tensor ($\alpha,\beta$=1, 2, 3) and $\Omega$ is the volume at the given
tensile strain. Tensile process along the [001], [110], and [111] directions
is implemented by increasing the lattice constants along these three
orientations step by step. At each step, the structure is fully relaxed until
all other five stress components vanished except that in the tensile direction.

\begin{figure}[ptb]
\includegraphics[width=0.8\textwidth]{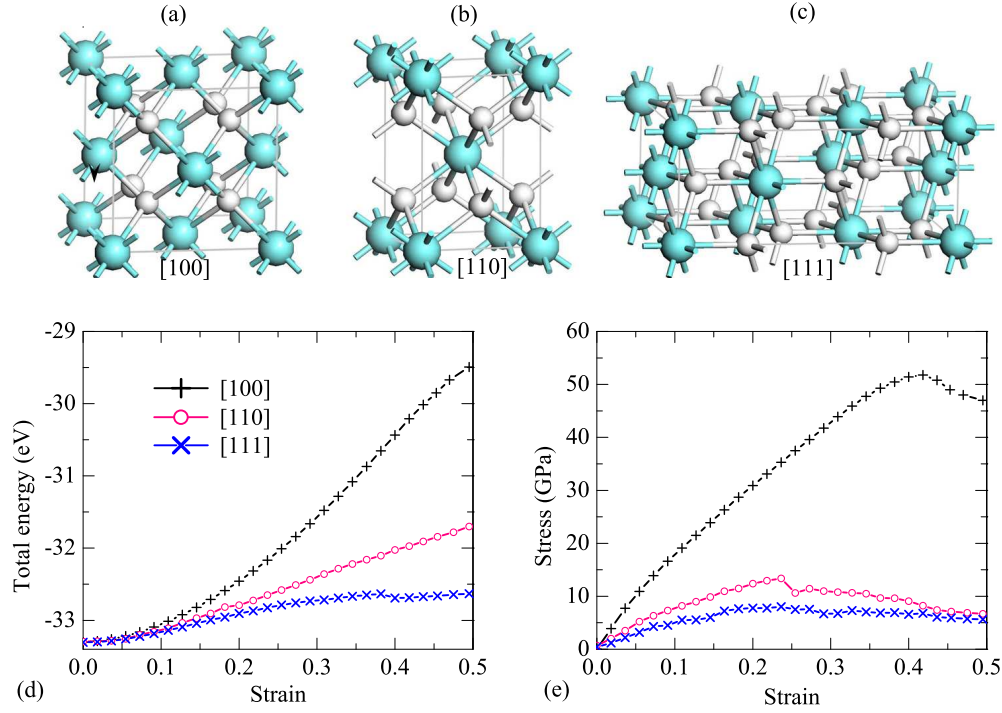}\caption{(Color online).
Schematic illustration of structure models for applying tensions along the
[100] (a), [110] (b), and [111] (c) directions. (d) and (e) Total energies per
formula unit and stresses for AmO$_{2}$ being pulled along the [100], [110],
and [111] directions.}%
\label{fig:fig3}%
\end{figure}

\begin{table}[ptb]
\caption{Calculated strain energy ($E_{strain}$) and stress maxima at the
critic points in the tensile processes.}%
\begin{ruledtabular}
\begin{tabular}{cccccccccccccccc}
& Strain & $E_{strain}$ & Stress \\
Direction & ~      & (eV/atom)    & (GPa)  \\
\hline
$[100]$ & 0.42 & 2.87 & 51.8 \\
$[110]$ & 0.24 & 0.64 & 13.4 \\
$[111]$ & 0.18 & 0.35 &  4.6 \\
\end{tabular}\label{estrain}
\end{ruledtabular}
\end{table}

The calculated total energies and stresses as functions of uniaxial tensile
strains for AmO$_{2}$ along the [100], [110], and [111] directions are shown
in Figs. 3(d) and 3(e). One can see that with increasing the tensile strains
along all the three directions, the energies increase with inflexions. By
differentiating the energy-strain curves, we find the inflexion points along
the [100], [110], and [111] directions to be the strain values of 0.42, 0.24,
and 0.18 respectively. At these strain values, the corresponding energy
increases are 2.87, 0.64, and 0.35 eV per atom in AmO$_{2}$, and the
corresponding stress maxima are 51.8, 13.4, and 7.6 GPa respectively, as shown
in Table IV. Since the stress-strain curves change from upward into downward
at these inflexion points, the stress maxima values are regarded as the
theoretical tensile strengths. Clearly along the three low-index directions,
the [100] direction has the largest, while the [111] direction has the
smallest tensile strength. This result is similar to what we have found in
another actinide dioxide PuO$_{2}$ \cite{Zhang10}. We can also understand this
result from chemical bonding analysis. Along the [100] direction, there are
eight Am-O bonds per formula unit for fluorite AmO$_{2}$, and the angle of all
eight bonds with respect to the pulling direction is 45$^{\circ}$. In
comparison, along the [110] and [111] directions only four and two Am-O bonds
make an angle of 45$^{\circ}$ with the pulling direction. The other four bonds
along the [110] direction are vertical to the pulling direction, while the
other six bonds make an angle of about 70.5$^{\circ}$ with the pulling
direction. It is evident that the bonds vertical to the pulling direction have
no contributions to the tensile strength, and the bonds parallel to the
pulling direction are easy to fracture under tensile deformation. Therefore,
the tensile strength along the [100] direction is stronger than that along
other two directions. Along all the three direction, no structural transition
has been observed under external strains in our present study. Besides, we
note that the stress in [110] direction experiences an abrupt decrease process
after strain up to 0.24. This is due to the fact that the corresponding four
Am-O bonds making an angle of 45$^{\circ}$ with the pulling direction have
been pulled to fracture under that strain. In comparison with PuO$_{2}$
\cite{Zhang10}, we can see that the theoretical tensile strengths of AmO$_{2}$
are much smaller along all the three low-index directions, indicating that
AmO$_{2}$ is easier to get deformed during tensile processes.

\subsection{Phonon dispersion curves and thermodynamic properties}

To our knowledge, no experimental or theoretical phonon frequency results have
been published for AmO$_{2}$. Here we carry out the first theoretical try in
obtaining its phonon dispersion curves, as well as thermodynamic properties.
The density functional perturbation theory (DFPT) and a 2$\times$2$\times$2
supercell containing 96 atoms is adopted to calculate the force constants of
AmO$_{2}$. Since our previously obtained phonon dispersion curves result by
using the same methods for PuO$_{2}$ accords moderately well with subsequent
experimental measurements, we are confident that our present results for
AmO$_{2}$ can be effective to certain degrees. The Born effective charges are
firstly calculated to reveal the dielectric properties of AmO$_{2}$, and to
modify the LO-TO (longitudinal optical and transverse optical branches of
phonon) splitting effects. For fluorite AmO$_{2}$, the effective charge
tensors for both Am and O are isotropic because of their position symmetry.
After calculation, the Born effective charges of Am and O ions are found to be
$Z_{\mathrm{Am}}^{\star}$ = 4.51 and $Z_{\mathrm{O}}^{\star}$ = $-$2.26,
respectively, within GGA+$U$ formalism with the choice of $U$=4.0 eV. We
present in the left panel of Fig. 4 (solid lines) our calculated phonon
dispersion curves for AmO$_{2}$. In the fluorite primitive cell, there are
three atoms (one Am and two O atoms). Therefore, nine branches of phonon
dispersion curves exist. One can see that the LO-TO splitting at $\Gamma$
point becomes evident by the inclusion of polarization effects. In addition,
due to the fact that americium atom is heavier than oxygen atom, the vibration
frequency of americium atom is lower than that of oxygen atom. As a
consequence, the phonon density of states for AmO$_{2}$ can be viewed as two
parts, as shown in the right panel of Fig. 4. One is the part lower than 4.0
THz where the main contribution comes from the americium sublattice while the
other part higher than 4.0 THz are dominated by the dynamics of the light
oxygen atoms.

\begin{figure}[ptb]
\includegraphics[width=0.6\textwidth]{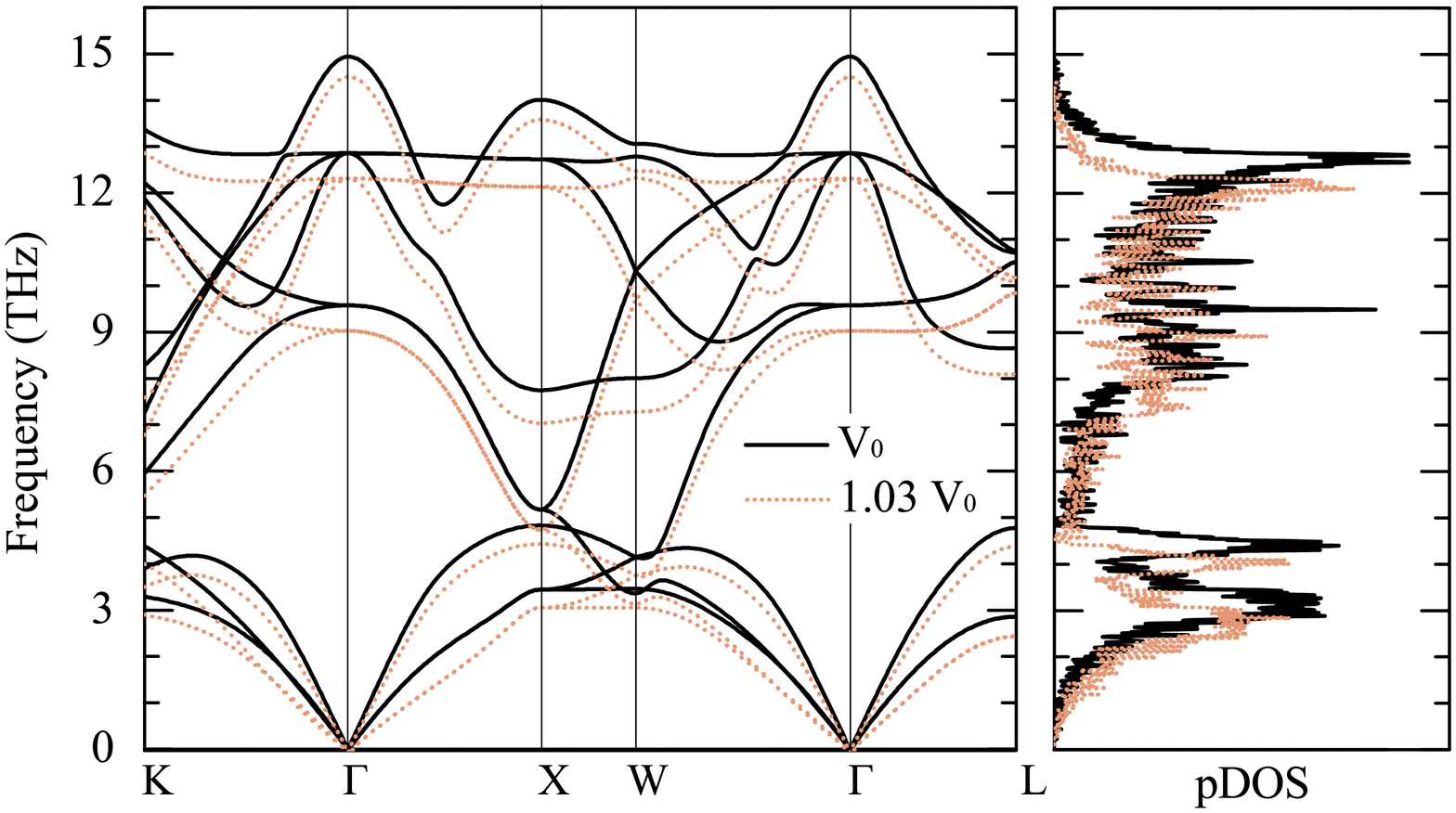}\caption{(Color online). Phonon
dispersion curves along the high symmetry lines of the Brillouin Zone (left
panel) and phonon density of states (right panel) for AmO$_{2}$ at the
equilibrium and expanded volumes. Solid and dashed lines represent for the
results at the equilibrium and expanded volumes respectively. The expansion
ratio is 0.03 for the expanded lattice calculations.}%
\label{fig:fig4}%
\end{figure}

To investigate the effects of strains on the phonon dispersion, we
also calculate the phonon spectrum for AmO$_{2}$ that is stretched
with an isotropic strain of 0.03, and show the results in Fig. 4
with dashed lines. Clearly after being expanded, both the acoustic
and optical branches of phonon experience frequency shift downward,
indicating the weakening of atomic interactions.

\begin{figure}[ptb]
\includegraphics[width=0.5\textwidth]{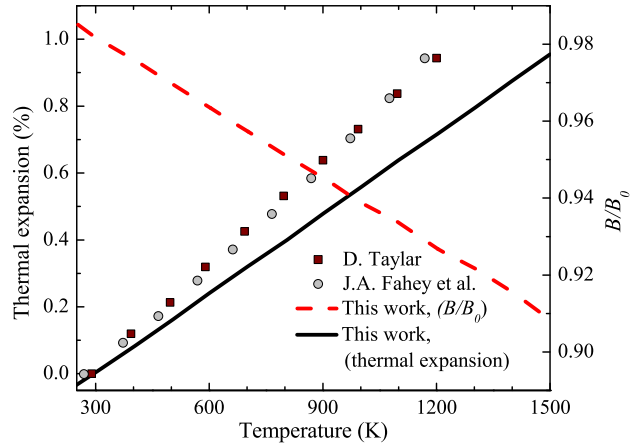}\caption{(Color online).
Temperature dependence of our calculated linear thermal expansion (solid line)
and bulk modulus ratio (dashed line), and the experimentally measured lattice
expansion results for AmO$_{2}$.}%
\label{fig:fig5}%
\end{figure}

The thermodynamic properties of a material are connected to its phonon
dispersion curves. By systematically calculating the Helmholtz free energies
at different lattice constants and different temperatures, we can determine
the lowest-energy lattice constant for AmO$_{2}$ at different temperatures. As
expected, the lattice constant enlarges as temperature increases. And the
obtained lattice expansion curve for AmO$_{2}$ is shown in Fig. 5, together
with the experimental results by Taylar \emph{et al.} \cite{Taylar85} and
Fahey \emph{et al.} \cite{Fahey74}. Similar linear relationships can be found
in our theoretical and previous experimental results. The small difference on
the increasing ratios comes from the fact that we only consider harmonic
vibrations during calculations of the phonon dispersion curves and the
Helmholtz free energy. By fitting the Brich-Murnaghan EOS for AmO$_{2}$ at
different temperatures for AmO$_{2}$, we also obtain the temperature
dependence of its bulk modulus $B$. And our result of $B/B_{0}$ as a function
of temperature $T$ is also shown in Fig. 5. One can see that as temperature
increases, the bulk modulus decreases, indicating that AmO$_{2}$ is easier to
get cracked at higher temperatures.

\begin{figure}[ptb]
\includegraphics[width=0.5\textwidth]{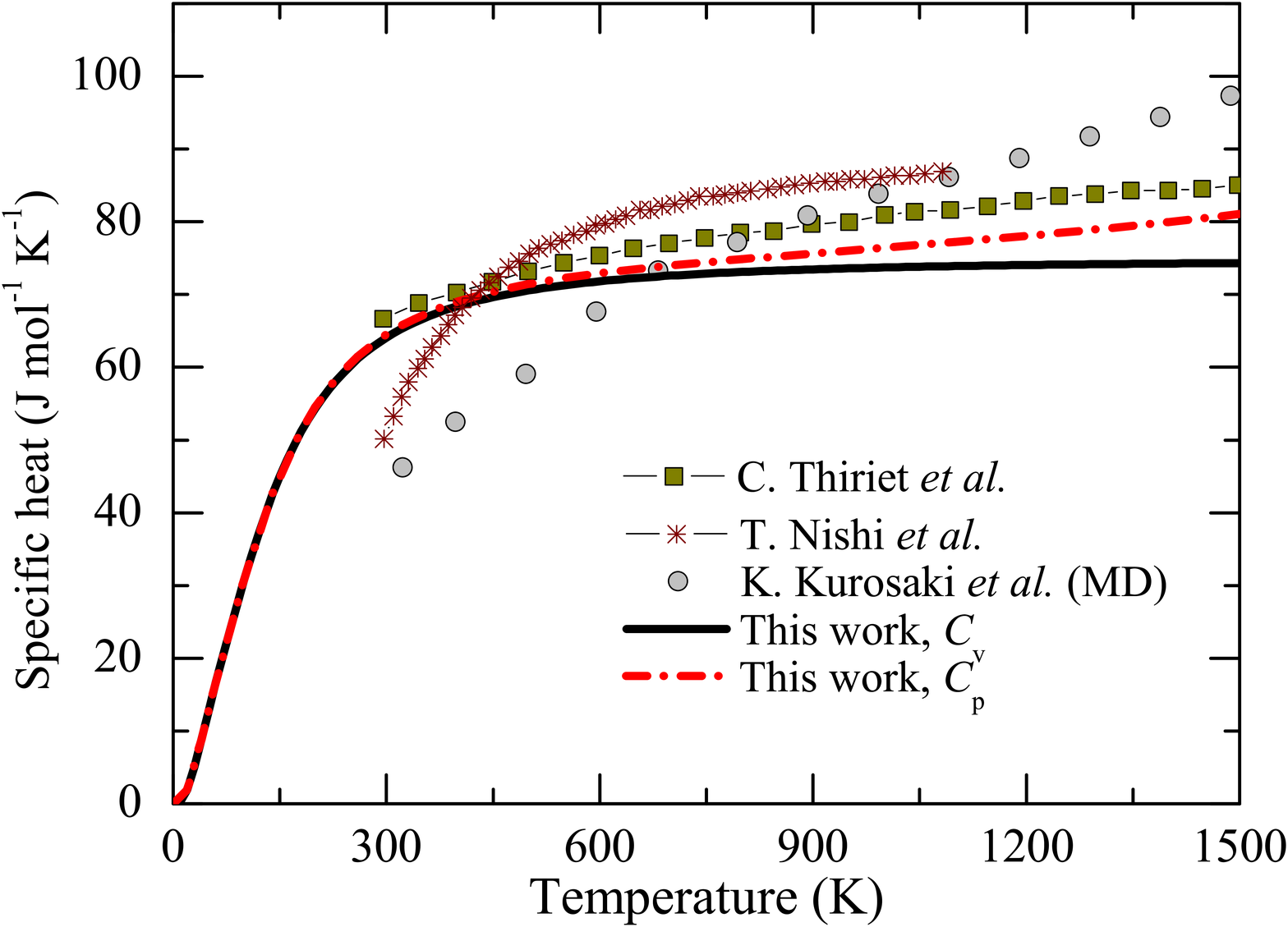}\caption{(Color online).
Calculated heat capacities of AmO$_{2}$ at constant volume ($C_{v}$) and
constant pressure ($C_{p}$), and corresponding experimental results as well as
other theoretical results.}%
\label{fig:fig6}%
\end{figure}

Within QHA the heat capacity at constant volumes can be calculated following
Eqs. (\ref{eq5}) and (\ref{eq6}). Besides, the heat capacity at constant
pressure $C_{p}$ can be further evaluated by using the relationship
\[
C_{p}-C_{v}=\alpha_{v}^{2}(T)B(T)V(T)T.
\]
We then calculate the heat capacities for AmO$_{2}$, and show the
results in Fig. 6. For comparison, the experimental data by Thiriet
\emph{et al.} \cite{Thiriet03} and Nishi \emph{et al.}
\cite{Nishi08} are also shown in Fig. 6. We can see that with
increasing the temperature, the value of $C_{p}$ increases
continuously, while the value of $C_{v}$ approaches to a constant of
3$R$ ($R$ is the gas constant). As clearly shown, our theoretical
result for $C_{p}$ is in good agreement with the experimental
measurements by Thiriet \emph{et al.} \cite{Thiriet03} at the
temperature range from 300 to 500 K. As temperature further
increases, the discrepancy between our result and the experimental
values \cite{Thiriet03,Nishi08} become larger. We conclude that the
disagreement of $C_{p}$ between theory and experiments in the high
temperature range mainly originates from the QHA we used. Due to
anharmonic effects, our obtained lattice expansion curve is lower
than experimental results, as shown in Fig. 5. The underestimation
on the lattice expansion of AmO$_{2}$ finally results in the
discrepancies between our calculated $C_{p}$ values and
corresponding experimental results through the above relationship
between $C_{p}$ and $C_{v}$.

\subsection{Gr\"{u}neisen parameters and thermal conductivity}

Based on the calculated phonon dispersion curves of AmO$_{2}$ at both the
equilibrium and expanded lattice volumes, we can further calculate the
Gr\"{u}neisen parameters for each vibration mode of AmO$_{2}$. Our obtained
Gr\"{u}neisen parameters at different $k$ points are shown in Fig. 7. Due to
the fact that after lattice expansion, atomic interactions are weakened and
phonon frequencies are lowered, the Gr\"{u}neisen parameters are positive for
almost all phonon branches. The only exception lies around the W point, where
the Gr\"{u}neisen parameter for the second acoustic branch of phonon has a
small negative value. This result comes from the coupling behavior of the two
lowest-frequency acoustic phonon branches at the W point. After lattice
expansion, the coupling disappears and the dispersion curves of these two
branches separate with each other, causing that the frequency of the higher
branch moves upward. As a result, the Gr\"{u}neisen parameters for this phonon
branch have small negative values around the W point.

We have explained above that the Umklapp scattering $\frac{1}{\tau_{U}}$ is
proportional to $\gamma^{2}$, thus, the Gr\"{u}neisen parameter provide an
estimate of the strength of anharmonicity in a compound. One can see from Fig.
7 that the two transverse acoustic branches of phonons have the largest
Gr\"{u}neisen parameters, indicating that these two branches are most
anharmonic and interact with the longitudinal acoustic mode strongly. In order
to describe quantitatively the anharmonic effects, we further calculate the
average Gr\"{u}neisen parameter ($\bar{\gamma}$) for the three acoustic
vibration modes by using the similar method proposed in Ref. \cite{Morelli02}:
$\bar{\gamma}=\sqrt{\langle\gamma_{i}^{2}\rangle}$. For AmO$_{2}$, the values
of $\bar{\gamma}_{TA}$, $\bar{\gamma}_{TA^{\prime}}$, $\bar{\gamma}_{LA}$ are
2.75, 2.48, and 2.16, respectively. And the average $\bar{\gamma}$ of the
three acoustic modes is then calculated to be 2.46. These transverse modes
play an important role in lattice thermal resistance. Furthermore, since the
velocity of transverse modes is much lower than the longitudinal one, the
lattice thermal conductivity is dominated by the lower-velocity transverse modes.

\begin{figure}[ptb]
\includegraphics[width=0.5\textwidth]{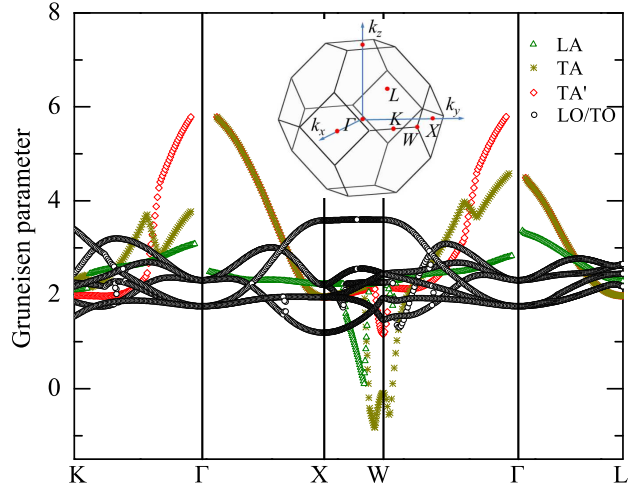}\caption{(Color online).
Calculated Gr\"{u}neisen parameters for AmO$_{2}$. Blue, red, and olive
symbols represent for the results corresponding to the two transverse acoustic
(TA and TA$^{\prime}$) and one longitudinal acoustic (LA) branches of phonon
respectively. The inset depicts the first Brillouin zone.}%
\label{fig:fig7}%
\end{figure}

\begin{figure}[ptb]
\includegraphics[width=0.5\textwidth]{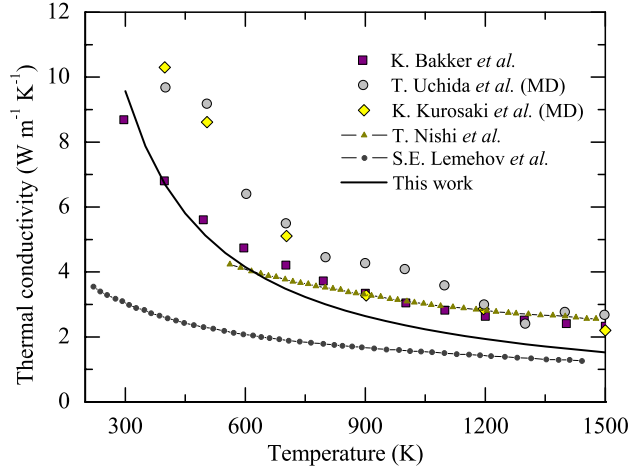}\caption{(Color online).
Calculated lattice thermal conductivity for AmO$_{2}$ at different
temperatures, and corresponding experimental as well as other theoretical
results.}%
\label{fig:fig8}%
\end{figure}

The acoustic Debye temperature closely related to the lattice thermal
conductivity of a material can be approximately calculated by using the
relationship \cite{Anderson59,Slack79}
\[
\Theta_{a}=\Theta n^{-1/3},
\]
where $\Theta$ and $n$ are the Debye temperature and number of atoms in the
unit cell, respectively. For AmO$_{2}$, $\Theta_{a}$ is calculated to be 233.0
K. Based on the relationship in Eq. (\ref{eq9}), and the above obtained
quantities, we calculate the lattice thermal conductivity for AmO$_{2}$ and
show the results from the temperature of 300 K to 1500 K in Fig. 8. For
comparison, previous experimental results by Bakker \textit{et al.}
\cite{Bakker98} and Nishi \textit{et al.} \cite{Nishi08}, as well as
theoretical results by Uchida \textit{et al.} \cite{Uchida09}, Kurosaki
\textit{et al.} \cite{Kurosaki04}, and Lemehov \textit{et al.}
\cite{Lemehov03} are also shown in Fig. 8. Clearly, in the temperature range
from 300 to 700 K, our theoretical results accord well with the experimental
measurements. At the higher temperature range of above 700 K, our calculated
thermal conductivity are a little smaller than the corresponding experimental
results. This disagreement might come from two different factors. Firstly, the
electronic thermal conductivity which occupies larger ratios at higher
temperatures is not considered here because of limitations of our
computational methods. Secondly, the Debye-Callaway model we used does not
take into account the optical branches of phonons, some of which have
appreciable group velocities and are able to contribute to thermal
conductivity. Overall, the coarse agreement between our theoretical results on
thermal conductivity with the experimental ones is fortuitous.

\section{CONCLUSIONS}

In summary, a thorough DFT+$U$ study has been performed to
investigate the electronic, mechanical, and thermodynamical
properties of AmO$_{2}$. It is found that the chemical bonding
between the americium and oxygen atoms show mixed ionic and covalent
characters. The valence band maximum and conduction band minimum are
contributed by 2$p$-5$f$ hybridized and 5$f$ electronic states
respectively. In comparison with PuO$_{2}$, the bonding length and
charge transfer in AmO$_{2}$ are both slightly smaller, indicating
stronger covalent and weaker ionic interactions in the latter than
than in the former. Through calculations of the elastic constants
and various moduli, we have predicted that compared to PuO$_{2}$,
AmO$_{2}$ is more stable against shear forces, but less stable
against pressures. The stress-strain relationship of AmO$_{2}$ has
also been examined along the three low-index directions by FPCTT
calculations. Our results show that the [100] and [111] directions
are the strongest and weakest tensile directions, respectively. In
comparison with PuO$_{2}$, the theoretical tensile strengths of
AmO$_{2}$ are smaller. At the thermodynamic aspects, the phonon
dispersion of AmO$_{2}$ has been calculated at different lattice
volumes, with the LO-TO splitting effect being modified through Born
effective charges. Subsequently, we have systematically obtained the
heat capacities and lattice expansion curve, both of which accord
well with the experiments. Finally, by differentiating the phonon
dispersion curves at different lattice constants, we have further
calculated the Gr\"{u}neisen parameters of AmO$_{2}$ and determined
its lattice thermal conductance at different temperatures, which
show the overall agreement with attainable experimental
measurements.

\begin{acknowledgments}
This work was supported by NSFC under Grant No. 51071032 and 11104170, and by
Foundations for Development of Science and Technology of China Academy of
Engineering Physics under Grants Nos. 2011A0301016 and 2011B0301060.
\end{acknowledgments}

\end{document}